# Recommendation Algorithms on Social Media: Unseen Drivers of Political Opinion

Waseq Billah

## Abstract

Social media broadly refers to digital platforms and applications that simulate social interactions online. This study investigates the impact of social media platforms and their algorithms on political interest among users. As social media usage continues to rise, platforms like Facebook and X (formerly Twitter) play increasingly pivotal roles in shaping political discourse. By employing statistical analyses on data collected from over 3,300 participants, this research identifies significant differences in how various social media platforms influence political interest. Findings reveal that moderate Facebook users demonstrate decreased political engagement, whereas even minimal engagement with X significantly boosts political interest. The study further identifies demographic variations, noting that males, older individuals, Black or African American users, those with higher incomes show greater political interest. The demographic analysis highlights that Republicans are particularly active on social media – potentially influencing their social media engagement patterns. However, the study acknowledges a crucial limitation: the lack of direct data regarding the content users are exposed to which is shaping their social media experiences. Future research should explore these influences and consider additional popular platforms to enhance the understanding of social media's political impact. Addressing these gaps can provide deeper insights into digital political mobilization, aiding policymakers, educators, and platform designers in fostering healthier democratic engagement.

# Introduction

Social media's rise since the early 2000s has been nothing short of spectacular. MySpace dominated the burgeoning industry initially, reaching an unprecedented one million users in 2004 (Ortiz-Ospina, 2019). It was rapidly outdone by the arrival of newcomers like YouTube and Facebook, each offering a fresh concept and swiftly dominating in its niche. To this day, Facebook stands out with a colossal global user base exceeding two billion, effectively encompassing more than one-third of the world's population (Ortiz-Ospina, 2019; Gottfried, 2024). Yet, the age of social media was only beginning.

Following these platforms, we have seen an ever-expanding ecosystem of social media services, each targeting a more specific audience. Instagram caters to visual storytelling, TikTok thrives on short-form, music-backed videos, Snapchat focuses on ephemeral communication, and LinkedIn serves professionals looking to network. The proliferation of these platforms has resulted in a drastic surge in the amount of time people spend online. In 2024, the average time spent on digital media was around 8 hours, which was almost twice as much as time spent on traditional media (Lee, 2024). With increasing time spent on social media, these platforms and their content have become an important presence in the lives of their users.

This steep rise in time spent on social media means that the content people consume, and the manner in which people interact with it, carries increasing weight in shaping their perceptions, opinions, and overall habits. Unfortunately, the content people are exposed to on social media are not universal but rather carefully curated to cater to the tastes of each individual user. While on the surface this may seem harmless (useful even), the far-reaching implications of this are profound.

This paper aims to explore whether the social media platforms and the algorithms they employ have an influence on the level of users' political interest. To do his, the paper begins with a brief overview of how recommendation algorithms emerged and how integrated they are with social media platforms. The literature on political interest and the factors that shape it are then discussed. To test the relationship between social media and political interest, statistical analyses are conducted using a secondary dataset and the findings as well as their implications are discussed.

## Background

The dawn of the commercial internet brought an explosion of online content and data. One of the first big strides came in the mid-1990s with the creation of "GroupLens" by researchers at the University of Minnesota (Resnick *et al.*, 1994). GroupLens introduced the idea of "collaborative filtering" - i.e. the ability to automate a very social process (asking friends for suggestions) at scale. However, as online catalogs grew and user bases ballooned, more sophisticated mathematical approaches were needed to handle huge datasets quickly.

Matrix factorization was one such leap forward (Koren, Bell and Volinsky, 2009). Although the name sounds technical, the fundamental principle is straightforward: you can arrange all the data about what users have liked (or clicked, bought, or rated) in a giant grid – a matrix. Each row represents a user, each column an item (like a movie or product), and the cells capture how much the user liked that item.

In parallel with matrix factorization, content-based filtering techniques gained prominence during the 2000s (Pazzani and Billsus, 2007). With content-based filtering, a system can, for instance, examine the plot, theme, or cast of a movie to make suggestions. If a user adores comedies starring a particular actor, the platform can suggest other comedies featuring that actor or a similar comedic style. This method also uses metadata – information about the item that is not the item itself – such as genre, director, or keywords.

In recent years, recommendation systems have undergone a profound transformation with the emergence of deep learning techniques. These cutting-edge algorithms, particularly neural networks, have revolutionized how recommendations are made (He *et al.*, 2017). Neural collaborative filtering models, a type of deep learning architecture specifically designed for recommendation tasks, have garnered significant attention for their ability to capture complex user-item interactions (Wu *et al.*, 2016). These models leverage neural networks to learn underlying representations of users and items, allowing them to discover intricate patterns and preferences that traditional methods may overlook. As we moved through history and followed the development of recommendation algorithms, there was almost a sense that something was missing. Browsing for recipes, online shopping, looking for music, etc. these were all well and good but not something that would really allow the algorithms to flex their muscles. For that the world needed social media to emerge.

Social media offers an endless supply of interactions: people like, share, comment, subscribe, follow, and post at any given moment, creating a tsunami of data. Not only do users consume posts, videos, and images, but they also actively create and distribute new content themselves, adding more data points for recommendation algorithms to process. Having gathered mountains of data from billions of users, these systems operate 24/7 to try and subliminally nudge our behavior (Törnberg and Uitermark, 2020). Whether its scrolling through our news feeds, discovering new connections, or joining communities, recommendation algorithms are at work behind the scenes, guiding us to content and experiences that it deems to be most relevant to us.

While personalized recommendations can enhance user engagement, there is also a huge risk of creating and perpetuating "filter bubbles" where users are only exposed to content that reinforces their existing beliefs and preferences (Pariser, 2012). If a platform's priority is to maximize time on site or user interaction, it might systematically feed users content that aligns with their existing viewpoints. The fact that social media does this is not disputed at all – but what we need to try and figure out is whether this exposure to algorithmically delivered content is having any influence on a user's political interests or perceptions.

## Related Works

Research on political interest has identified numerous factors contributing to individuals' engagement and attitudes towards politics. Among these, social media has emerged as a particularly influential factor. Kubin and von Sikorski (2021) demonstrated that social media platforms significantly contribute to political polarization, largely due to algorithms that curate content aligned with users' pre-existing beliefs, effectively creating ideological echo chambers. This phenomenon intensifies political divides by limiting exposure to diverse perspectives, thereby reinforcing existing opinions and biases.

Further emphasizing the centrality of social media, Walker and Matsa (2021) highlighted that these platforms, particularly Twitter (now X), have become primary sources of news for many individuals. The instantaneous and interactive nature of platforms like X facilitates rapid dissemination of news and political information, significantly influencing public discourse and opinion formation. Political affiliation significantly influences attitudes towards information credibility and political interest (Rhodes, 2022). In particular, Republicans have shown higher levels of suspicion regarding fact-checking and mainstream media sources (Allcott and Gentzkow, 2017). This skepticism can influence their engagement patterns on social media and their responsiveness to political content, shaping overall political interest and behavior.

Studies also show that demographic characteristics can substantially shape political interest. Males are typically more vocal and expressive regarding political interests online, often engaging more actively in political discourse (Koc-Michalska *et al.*, 2021). This heightened visibility aligns with findings by Bos et al. (2022), suggesting that men generally possess greater political aspirations, potentially driving higher levels of engagement and interest.

Age also proves to be an essential determinant of political interest. Older populations consistently exhibit greater political interest compared to younger demographics (Bos *et al.*, 2022; Rhodes, 2022; Zhang, 2022). This increased engagement among older individuals (especially in the case of males) could be attributed to various factors, including life stage, experience, and heightened awareness of the political system's direct impacts on their lives.

Race and ethnicity further influence political interest, with White Americans generally reporting higher levels of political interest (Carnes, 2018; Harris and Rivera-Burgos, 2021). This is primarily due to various structural and historical factors, including access to resources, representation, and perceived political efficacy, might contribute to differing levels of political engagement across racial groups. Similarly, income levels have also been consistently linked to political engagement (Carnes, 2018). Individuals in higher income brackets demonstrate greater political interest and engagement, potentially due to their higher stake in policies affecting taxation, economic regulation, and social welfare programs. Economic resources and educational opportunities likely facilitate easier access to political information and participation. As evidenced by the literature, an analysis of political interest would not be complete without incorporating these factors into our model. The following sections outline the data and methods used for this study.

## Data and Methods

The dataset used for this study can be found in the OpenICPSR repository under the University of Michigan (Rhodes, 2022). The data was collected using an online survey on over 3,300 participants. 53% of the participants were male and the ages of participants ranged from 19 to 99 – with the average age being 37.7. The majority of participants were white (66%), followed by Black or African American (18%); Hispanic or Latino (6%); Asina or Pacific Islander (6%); Native American (3% and other (1%). There was a good mixture of participants from different party affiliations and income groups (see Appendix).

### *Variables*

The primary dependent variable is political interest. The dataset includes an ordinal variable for this (1 = very interested – 4 = not at all interested in politics). I recoded this so that higher coded categories reflect increasing interest in politics (i.e. 1 = not at all interested – 4 = very interested) for easier analysis.

As for independent variables of interest, this study looked at social media usage for different platforms. This is included in the dataset in the form of ordinal variables that record the number of days spent on social media (ranging from none to 7 days). While, there is no direct data on the algorithms or the types of content the users are being exposed to, this study assumes that users of the major platforms will have been exposed to algorithmic nudges by just being active in the platforms and so will look to see if this correlates with their interest in politics.

In terms of platforms, this study focused on the data from four of the most popular social media platforms: Facebook, X, Instagram and Snapchat. Other variables included in the analyses are age, race, gender, party affiliation and income level as these are factors known to affect political interest and participation according to the literature.

*Data Analysis*

The dependent variable is ordinal – so a form of ordinal logistic regression needed to be used. Within ordinal logistic regression, there is an important condition called the "parallel odds" assumption. This suggests that the way a predictor influences moving from the lowest category to the next is the same as it influences moving from, say, the second category to the third, and so on. Since the Brant test revealed that the parallel odds assumption was violated (see Appendix), the stereotype logistic regression was used (Anderson, 1984). This is designed to handle situations where the parallel odds assumption does not hold, offering a more flexible way to model the relationships between predictors and an ordinal dependent variable. In terms of interpretations, a positive coefficient will indicate that participants were more likely to fall in higher coded categories of the dependent variable (in this case - be more interested in politics) while a negative coefficient will indicate that participants were less likely to fall in higher coded categories (i.e. be less interested in politics). Robust standard errors were used to ensure the model is in its most efficient form.

The Chi-square test was also carried out to see if there were any significant differences between racial groups and supporters of different political parties when it came to social media usage. In these cases, the p-values and the effect size was reported in the form of Cramer's *V* and described using the classifications outlined by Cohen and Vaske (Cohen, 1988; Vaske, 2019). All statistical analyses were conducted using STATA.

# Results

This section presents the results derived from regression analyses and chi-square tests exploring the influence of social media platforms and demographic factors on political interest. The findings highlight distinct patterns in social media usage and its associated impacts across different groups of participants.

*Table 1: Stereotype Logistic regression – effect of predictors on political interest*

| VARIABLES | Interest in Politics |
|---|---|
| **Facebook Usage *(Baseline - None)*** | |
| One Day | -.875** |
| Two Days | -1.457*** |
| Three Days | -.926** |
| Four Days | -1.014** |
| Five Days | -.293 |
| Six Days | -.201 |
| Seven Days | -.205 |
| **X Usage *(Baseline - None)*** | |
| One Day | .819** |
| Two Days | 1.591*** |
| Three Days | 1.413*** |
| Four Days | 1.274*** |
| Five Days | 1.784*** |
| Six Days | 2.266*** |
| Seven Days | 2.280*** |
| **Instagram Usage *(Baseline - None)*** | |
| One Day | .444 |
| Two Days | .562 |
| Three Days | -.588 |
| Four Days | -.514 |
| Five Days | .052 |
| Six Days | -.348 |
| Seven Days | -.329 |
| **Snapchat Usage *(Baseline - None)*** | |
| One Day | -.067 |
| Two Days | -.162 |
| Three Days | -.368 |
| Four Days | -.152 |
| Five Days | .028 |
| Six Days | -.008 |
| Seven Days | -.248 |
| Gender *(Baseline - Female)* | |
| Male | .526*** |
| Age | .040*** |
| Race *(Baseline – White)* | |
| Hispanic or Latino | -.650* |
| Black of African American | .431* |
| Native American or American Indian | -.205 |
| Asian or Pacific Islander | -1.258*** |
| Other | -.304 |
| Party Affiliation *(Baseline – No Party)* | |
| Democrat | 4.114*** |
| Republican | 3.911*** |
| Independent | 2.975*** |
| Other Party | 4.422*** |
| Income Level *(Baseline – Less than $10,000)* | |
| $10,000 to less than $20,000 | .667 |
| $20,000 to less than $30,000 | .947* |
| $30,000 to less than $40,000 | 1.032* |
| $40,000 to less than $50,000 | 1.321** |
| $50,000 to less than $75,000 | 1.207** |
| $75,000 to less than $100,000 | 1.105** |
| $100,000 to less than $150,000 | .933* |
| $150,000 or more | 1.244* |

*$p < 0.05$; ** $p < 0.01$; *** $p < 0.001$

The regression analysis indicates that, among the four social media platforms examined, Facebook and X significantly influenced participants' political interest, albeit in contrasting ways. For Facebook, participants who spent four days or fewer on the platform were, on average, significantly less likely to be interested in politics compared to those who did not use Facebook at all. By contrast, X had a stronger positive influence: participants who used X at least one day per week were, on average, significantly more likely to be interested in politics than those who did not use X.

In terms of demographic characteristics, the findings align with previous research. Men were significantly more likely than women to be interested in politics, and older participants were more likely than younger participants to exhibit a high level of political interest.

The result for race in Table 1 is also quite interesting as compared to White Americans, Hispanic and Asian populations are less likely to be interested in politics but black or African Americans are more likely to be interested in politics. As for party affiliation and income levels, the results were as expected – users identifying with a political party were more likely to be interested in politics and everyone aside from the poorest Americans were more likely to be interested in politics.

*Table 2: Social media usage amongst different races*

| | White (%) | Hispanic or Latino (%) | Black or African American (%) | Native American or American Indian (%) | Asian or Pacific Islander (%) | Other (%) | Chi-Squared ($\chi^2$) Value | *p*-value | Cramer's *V* |
|---|---|---|---|---|---|---|---|---|---|
| Facebook Usage (Days) | | | | | | | | | |
| None | 14 | 18 | 8 | 0 | 12 | 12 | | | |
| One | 8 | 11 | 11 | 15 | 15 | 9 | | | |
| Two | 6 | 5 | 5 | 5 | 7 | 6 | | | |
| Three | 6 | 7 | 6 | 11 | 8 | 6 | 170.249 | .000*** | .101 |
| Four | 6 | 6 | 7 | 8 | 4 | 6 | | | |
| Five | 9 | 11 | 16 | 11 | 7 | 10 | | | |
| Six | 9 | 7 | 19 | 15 | 5 | 11 | | | |
| Seven | 42 | 35 | 29 | 35 | 42 | 33 | | | |
| X Usage (Days) | | | | | | | | | |
| None | 32 | 29 | 12 | 7 | 34 | 54 | | | |
| One | 10 | 8 | 9 | 17 | 11 | 15 | | | |
| Two | 7 | 7 | 6 | 8 | 9 | 4 | 198.051 | .000*** | .109 |
| Three | 8 | 10 | 9 | 7 | 8 | 4 | | | |
| Four | 9 | 8 | 10 | 15 | 6 | 6 | | | |
| Five | 9 | 9 | 16 | 13 | 9 | 2 | | | |

| | | | | | |
|---|---|---|---|---|---|
| Six | 8 | 8 | 17 | 19 | 7 | 2 |
| Seven | 18 | 20 | 21 | 13 | 16 | 13 |

*$p < 0.05$; **$p < 0.01$; ***$p < 0.001$

Table 2 takes a deeper dive into the social media usage patterns amongst different races – especially in relation to Facebook and X. It is evident that time spent on social media varies significantly amongst different races with a small or minimal effect (Cohen, 1988; Vaske, 2019). Interestingly, Black or African Americans were very active on social media – with 71% and 64% of them spending 4 or more days on Facebook and Twitter respectively. This was higher than any of the other racial groups.

*Table 3: Social media usage and party affiliation*

| | No Preference (%) | Democrat (%) | Republican (%) | Independent (%) | Other Party (%) | Chi-Squared ($\chi^2$) Value | *p*-value | Cramer's V |
|---|---|---|---|---|---|---|---|---|
| **Facebook Usage (Days)** | | | | | | | | |
| None | 25 | 12 | 6 | 21 | 44 | | | |
| One | 9 | 10 | 10 | 9 | 3 | | | |
| Two | 5 | 6 | 6 | 5 | 6 | | | |
| Three | 5 | 6 | 7 | 7 | 0 | 177.419 | .000*** | .116 |
| Four | 5 | 6 | 8 | 5 | 0 | | | |
| Five | 9 | 9 | 12 | 7 | 6 | | | |
| Six | 4 | 11 | 13 | 8 | 8 | | | |
| Seven | 39 | 40 | 38 | 39 | 33 | | | |
| **X Usage (Days)** | | | | | | | | |
| None | 62 | 26 | 20 | 40 | 39 | | | |
| One | 9 | 10 | 9 | 11 | 19 | | | |
| Two | 5 | 7 | 7 | 7 | 0 | | | |
| Three | 7 | 8 | 8 | 6 | 6 | 209.671 | .000*** | .126 |
| Four | 5 | 9 | 11 | 6 | 0 | | | |
| Five | 1 | 11 | 12 | 6 | 3 | | | |
| Six | 2 | 9 | 13 | 6 | 3 | | | |
| Seven | 10 | 20 | 19 | 15 | 31 | | | |

*$p < 0.05$; **$p < 0.01$; ***$p < 0.001$

Finally, Table 3 presents the results of a chi-square test on social media usage by political party affiliation. On average, participants identifying as Republicans reported significantly higher engagement compared to other participants with 72% of them using Facebook four or more days per week, and 55% of them using X four or more days per week.

## Discussion

The findings from this study strongly suggest that social media platforms and the type of content that users encounter on these platforms can significantly shape political interests of users. This is evident from the regression analysis, which revealed clear impacts on political interest stemming specifically from the use of Facebook and X. Facebook users who engaged with the platform moderately (four days or fewer per week) demonstrated lower political interest. This could suggest that minimal or intermittent Facebook use might dilute political interest or, alternatively, that casual users tend to use Facebook more for social or recreational purposes rather than for active political engagement. In contrast, X exhibited a considerably broader influence. Even minimal engagement – spending just one day a week – significantly increased the likelihood of a user's interest in politics.

The racial dimension observed in the usage patterns further highlights the intersection of social media use and political interest. As illustrated in Tables 1 and 2, Black or African American participants were particularly active on social media platforms and also was recorded as being more interested in politics compared to other racial groups. This relationship suggests that active engagement with X might play a pivotal role in enhancing political interest within this community. Given the platform's strength in mobilizing social movements and providing a voice to historically marginalized groups, this result aligns with broader observations regarding X's role in empowering political discourse among minority populations.

Additionally, political party affiliation emerged as another critical factor influencing social media use. The findings from Table 3 indicate that participants identifying as Republican were considerably more active on social media. This substantial social media presence among Republican voters underscores the platform's potential for political mobilization and opinion shaping within partisan communities. The higher engagement level among Republican users might reflect deliberate outreach strategies by political entities or could indicate stronger alignment between Republican political messaging and content circulated on these platforms.

While the dataset was compiled a couple years ago, the results line up well in current events - thus further reinforcing the significance of these findings (Cousens, 2024). In recent years, voter demographics and behaviors have noticeably shifted, with social media increasingly being recognized as a significant factor driving these changes. Elections, particularly in the United States and other democracies, have witnessed remarkable transformations in how voter bases form opinions, mobilize support, and engage politically, largely facilitated by social media channels. Platforms like X and Facebook have not only become essential tools for political campaigns but also serve as critical arenas for public political discourse and debate. These platforms provide candidates and parties unprecedented reach and immediacy, shaping political narratives in real time.

Given these developments, an essential avenue for future research is exploring how users interact with content and how social media algorithms subsequently influence content visibility and user engagement. The algorithmic curation of content presents a fascinating and significant area for further study. Specifically, understanding how algorithms shape users' newsfeeds, prioritize certain political narratives, and create echo chambers or polarized communities would yield critical insights. Such insights could inform policymakers, platform developers, and civil society actors aiming to mitigate misinformation, promote balanced political discourse, and enhance democratic engagement.

Moreover, future research could delve deeper into analyzing the types of political narratives proliferating on social media platforms and their specific roles in shaping political attitudes and behaviors. Examining narrative strategies, including the framing of issues, characterization of political actors, and emotional appeals, could offer valuable perspectives on how narratives effectively mobilize or demobilize specific voter groups.

## Conclusion

This study highlights that platforms like Facebook and X play a critical role in shaping political interest and behavior. However, while the dataset is sufficiently large, it lacks detailed information about the algorithms that govern content visibility, and the exact nature of messages users are encountering on these platforms. Understanding these aspects is crucial because algorithms significantly determine what content is prioritized and presented, thereby shaping users' political perceptions and influencing their engagement patterns. Without this knowledge, the study cannot fully capture the mechanisms through which social media platforms exert influence.

Future research should therefore explicitly investigate the algorithms responsible for content curation on social media platforms and their subsequent impacts on political behavior. Detailed analyses of content exposure – such as the types of political narratives promoted, frequency of political messaging, and emotional and persuasive tactics employed – would substantially enhance our understanding of social media's role in shaping public discourse.

Additionally, including data from increasingly popular social media platforms, particularly TikTok, would provide critical insights. TikTok's rapid growth, especially among younger demographics, positions it uniquely within the digital political landscape. Its algorithmic model, characterized by highly personalized, short-form video content, may yield distinct patterns of political influence compared to other platforms. Overall, expanding research to address these limitations and incorporate newer platforms would significantly enrich our comprehension of social media's evolving role in contemporary politics, helping inform effective communication strategies and policy interventions aimed at fostering healthier democratic dialogue and participation.

# Appendix

Brant test results

```
. brant

Brant test of parallel regression assumption

                  |    chi2     p>chi2     df
            ------+------------------------------
              All |   54.19     0.000      18
            ------+------------------------------
        socialm2_1|    4.76     0.092       2
        socialm2_3|   13.37     0.001       2
        socialm2_2|    7.14     0.028       2
        socialm2_5|    3.29     0.193       2
             race |    0.03     0.986       2
            party |   17.14     0.000       2
          income2 |    2.94     0.229       2
             male |    3.30     0.192       2
              age |    0.78     0.678       2

A significant test statistic provides evidence that
regression assumption has been violated.
```

Demographics Data

| Demographics | % |
|---|---|
| Gender | |
| Male | 53 |
| Female | 47 |
| Race | |
| White | 66 |
| Hispanic or Latino | 6 |
| Black or African American | 18 |
| Native American or American Indian | 3 |
| Asian or Pacific Islander | 6 |
| Other | 1 |
| Party Affiliation | |
| No Party | 3 |
| Democrat | 39 |
| Republican | 36 |
| Independent | 20 |
| Other | 1 |
| Income Level | |
| Less than $10,000 | 5 |
| $10,000 to less than $20,000 | 9 |
| $20,000 to less than $30,000 | 11 |
| $30,000 to less than $40,000 | 13 |
| $40,000 to less than $50,000 | 14 |
| $50,000 to less than $75,000 | 24 |
| $75,000 to less than $100,000 | 14 |
| $100,000 to less than $150,000 | 8 |
| $150,000 or more | 3 |